\PassOptionsToPackage{unicode=true}{hyperref} 

\documentclass[runningheads]{ceurart}
\usepackage[T1]{fontenc}
\usepackage[utf8]{inputenc}

\usepackage{hyperref}
\hypersetup{
            pdftitle={A Curiously Effective Backtracking Strategy for Connection Tableaux},
            pdfkeywords={connection tableaux, backtracking, exclusive
cut, REX, leanCoP, meanCoP},
            pdfborder={0 0 0},
            breaklinks=true}

\providecommand{\tightlist}{}

\usepackage{amsmath}
\usepackage{amsthm}

\usepackage{xcolor}

\usepackage{tikz}
\usepackage{forest}
\usepackage{todonotes}

\usetikzlibrary{arrows,calc,positioning,tikzmark}
\tikzset{
    >=stealth,
}

\usepackage{pgfplots}
\pgfplotsset{compat=1.9}
\usepgfplotslibrary{units}

\usepackage{subcaption}

\usepackage{booktabs}
\usepackage{calc}
\usepackage{makecell}

\usepackage{prftree}

\usepackage{stmaryrd}
\usepackage{pifont}

\newcommand{\Lightning}{$\lightning$}

\makeatletter
\def\cleartheorem#1{%
    \expandafter\let\csname#1\endcsname\relax
    \expandafter\let\csname c@#1\endcsname\relax
}
\makeatother

\newtheorem{definition}{Definition}
\newtheorem{example}{Example}

\usepackage{centerfloat}
\usepackage{includetexgraphics}

\renewcommand\textbf[1]{\fontseries{b}\selectfont{#1}}

\begin{document}
\title{A Curiously Effective Backtracking Strategy for Connection
Tableaux}

\author{Michael
Färber}[orcid=0000-0003-1634-9525,email=michael.faerber@gedenkt.at]
\date{}


\conference{AReCCa 2023: Automated Reasoning with Connection Calculi, 18
September 2023, Prague, Czech Republic}
\copyrightyear{2023}
\copyrightclause{Copyright for this paper by its authors. Use permitted
under Creative Commons License Attribution 4.0 International (CC BY
4.0).}

\maketitle
\begin{abstract}
Automated proof search with connection tableaux, such as implemented by
Otten's leanCoP prover, depends on backtracking for completeness.
Otten's restricted backtracking strategy loses completeness, yet for
many problems, it significantly reduces the time required to find a
proof. I introduce a new, less restricted backtracking strategy based on
the notion of \emph{exclusive cuts}. I implement the strategy in a new
prover called \emph{meanCoP} and show that it greatly improves upon the
previous best strategy in leanCoP.
\end{abstract}

\hypertarget{introduction}{%
\section{Introduction}\label{introduction}}

Bibel's connection method \citep{DBLP:books/lib/Bibel87a} is a proof
search method similar to Andrews's matings
\citep{DBLP:journals/jacm/Andrews81}. Compared to other proof search
methods such as resolution, the connection method has several merits: It
is goal-oriented, enabling natural conjecture-directed proof search. It
can be used with relatively little effort for non-classical logics such
as intuitionistic or modal logics
\citep{DBLP:conf/cade/Otten08, DBLP:conf/cade/Otten14}, and non-clausal
search \citep{DBLP:conf/ijcai/Otten17}. Finally, most connection calculi
have only very few and simple rules, making it easy to certify proofs in
proof assistants such as HOL Light
\citep{DBLP:conf/cpp/KaliszykUV15, DBLP:conf/tableaux/0002K19}.

One of the most influential connection provers is Otten's leanCoP
\citep{DBLP:conf/cade/Otten08}. Its outstanding ratio between code size
and effectiveness has made it a frequently used vehicle to experiment
with new search strategies. leanCoP uses bounded depth-first search
together with iterative deepening to explore larger and larger potential
proofs. As the proof search is not confluent, leanCoP employs
backtracking to preserve completeness.

This article studies backtracking that guides connection proof search.
In particular, a backtracking strategy deals with the question: When a
literal \(L\) is solved with a proof \(P\), which alternative proofs
\(P'\) to solve \(L\) can proof search consider afterwards? A complete
backtracking strategy does not impose any restriction on \(P'\); that
is, it allows proof search to consider all different proofs \(P'\) for
\(L\). Otten showed that by restricting backtracking, the prover becomes
significantly more effective for many problems as this reduces the
search space, at the expense of losing completeness
(\autoref{backtracking}). His ``restricted backtracking'' strategy
prevents exploring any alternative proof \(P'\) for a solved literal
\(L\). In this paper, I introduce a novel incomplete strategy called
``less restricted backtracking'', which prevents exploring any
alternative proof \(P'\) for a solved literal \(L\) where \(P'\) starts
with the same root step as \(P\) (\autoref{less-restricted}). In other
words, for any literal \(L\), unrestricted backtracking considers all
proofs, restricted backtracking considers only a single proof, and less
restricted backtracking considers only proofs with differing root steps.

I unexpectedly discovered less restricted backtracking upon implementing
a new prover called \emph{meanCoP} based on leanCoP
(\autoref{implementation}). The new strategy improves upon the former
best strategy in leanCoP dramatically (\autoref{evaluation}).

\hypertarget{preliminaries}{%
\section{Preliminaries}\label{preliminaries}}

In this article, we will use classical first-order logic without
equality. However, the techniques shown can be also applied to
non-classical logics.

A term \(t\) is either a variable (denoted by \(x\), \(y\), \(z\)) or
the application of a constant (denoted by \(\mathsf{a}\),
\(\mathsf{b}\), \(\mathsf{c}\)) to terms. An atom \(A\) is the
application of a predicate (denoted by \(\mathsf{p}\), \(\mathsf{q}\),
\(\mathsf{r}\)) to terms. Predicates and constants have associated fixed
arities. A literal \(L\) is an atom \(A\) or its negation \(\lnot A\).
The complement \(\overline{\phantom{A}}\) of a literal is defined such
that \(\overline{A} = \lnot A\) and \(\overline{\lnot A} = A\). A term
substitution \(\sigma\) is a mapping from variables to terms. Applying a
substitution \(\sigma\) to a literal \(L\), denoted as \(\sigma L\),
substitutes all variables of \(L\) with their mappings. Two literals
\(L_1\), \(L_2\) can be unified under a substitution \(\sigma\) if
\(\sigma L_1 = \sigma L_2\).

A formula in conjunctive normal form (CNF) is a conjunction (\(\land\))
of disjunctions (\(\lor\)) of literals. A clause is a set of literals,
and a matrix is a set of clauses. We interpret a clause as the
disjunction of its literals, and we interpret a matrix as the
conjunction of its (interpreted) clauses. It is easy to see that for
each formula in CNF, there is an equivalent matrix.

\begin{example}\label{ex:matrix}Consider the formula
\[(\mathsf{p}(x) \lor \mathsf{q}(x)) \land
  (\lnot \mathsf{p}(y) \lor \mathsf{r}(y)) \land
  \lnot \mathsf{p}(z) \land
  \lnot \mathsf{r}(\mathsf{a}) \land
  \lnot \mathsf{r}(\mathsf{b}) \land
  \lnot \mathsf{q}(\mathsf{c}).\] Its equivalent matrix is \[M = \left[
  {\left[\begin{matrix}\mathsf{p}(x) \\ \mathsf{q}(x)\end{matrix}\right]}
  {\left[\begin{matrix}\lnot \mathsf{p}(y) \\ \mathsf{r}(y)\end{matrix}\right]}
  {\left[\begin{matrix}\lnot \mathsf{p}(z)\end{matrix}\right]}
  {\left[\begin{matrix}\lnot \mathsf{r}(\mathsf{a})\end{matrix}\right]}
  {\left[\begin{matrix}\lnot \mathsf{r}(\mathsf{b})\end{matrix}\right]}
  {\left[\begin{matrix}\lnot \mathsf{q}(\mathsf{c})\end{matrix}\right]}
  \right],\] which we will use as running example throughout this
paper.\end{example}

In this paper, we treat proof search using the clausal connection
tableaux calculus
\citep{DBLP:books/el/RV01/LetzS01, DBLP:journals/jsc/OttenB03}.\footnote{Unlike
  this article,
  \citep{DBLP:books/el/RV01/LetzS01, DBLP:journals/jsc/OttenB03} use
  disjunctive normal form (DNF) and check for validity of a formula. The
  two presentations are dual; in particular, the DNF of a formula is
  valid iff the CNF of its negation is unsatisfiable.}

\begin{definition}[Connection Calculus]The axiom and the rules of the
clausal connection calculus are given in \autoref{fig:clausal-calculus}.
The words of the connection calculus are tuples
\(\langle C, M, Path \rangle\), where \(M\) is a matrix, and \(C\) and
\(Path\) are sets of literals or \(\varepsilon\). \(C\) is called the
\emph{subgoal clause} and \(Path\) is called the \emph{active path}. In
the calculus rules, \(\sigma\) is a global (or rigid) term substitution;
that is, it is applied to the whole derivation.\end{definition}

\begin{figure}
\includegraphics{prftree/clausal-calculus.tex}
\caption{Clausal connection calculus rules.}
\label{fig:clausal-calculus}
\end{figure}

\sloppy

An application of a proof rule is called a proof step. A derivation for
\(\langle C, M, Path \rangle\) with the term substitution \(\sigma\), in
which all leaves are axioms, is called a \emph{connection proof} for
\(\langle C, M, Path \rangle\). A connection proof for
\(\langle \varepsilon, M, \varepsilon \rangle\) is called a connection
proof for \(M\).

Bibel proved soundness and completeness of the calculus: for any formula
\(F\) in CNF, we have that \(F\) is unsatisfiable iff there is a
connection proof for the matrix corresponding to \(F\)
\citep{DBLP:books/lib/Bibel87a}.

Proof search proceeds by constructing derivations from bottom to top. We
can understand a derivation for \(\langle C, M, Path \rangle\) as an
attempt to prove \(\left(M \land Path\right) \implies C\), where we
interpret \(Path\) as conjunction of its literals. By interpreting
\(\varepsilon\) as empty set, a derivation for
\(\langle \varepsilon, M, \varepsilon \rangle\) can be seen as a proof
attempt of \(M \implies \bot\).

We say that any reduction or extension step as in
\autoref{fig:clausal-calculus} \emph{connects} \(L\) to \(L'\). We
illustrate this by drawing an arrow from \(L\) to \(L'\) in the matrix.
In this paper, we will only use extension steps in examples.

Let us walk through a \emph{failed} proof search attempt for the matrix
\(M\) from \autoref{ex:matrix}, and show its graphical representation as
well as its resulting derivation in the calculus.

\begin{figure}
\includegraphics{tikz/search.tex}
\caption{Graphical representation of proof search.}
\label{fig:search}
\end{figure}

\begin{example}\label{ex:failed}Consider matrix \(M\) from
\autoref{ex:matrix}. Matrix (1) of \autoref{fig:search} illustrates a
proof search attempt through \(M\). We write the proof step \(n\) in
matrix \(m\) as (\(m\).\(n\)) and mark situations in which we are stuck
with \Lightning. The proof search proceeds as follows: We first choose
the first clause in \(M\) as start clause. This obliges us to connect
both \(\mathsf{p}(x)\) and \(\mathsf{q}(x)\). We start with
\(\mathsf{p}(x)\), which we choose to connect in step (1.1) to
\(\lnot \mathsf{p}(y)\), setting \(\sigma(x) = y\). This in turn obliges
us to connect \(\mathsf{r}(y)\), which we choose to connect in step
(1.2) to \(\lnot \mathsf{r}(\mathsf{a})\), setting
\(\sigma(x) = \sigma(y) = \mathsf{a}\). We are now left with the
obligation to connect \(\mathsf{q}(x)\). However, at this point (1.3),
we cannot connect \(\mathsf{q}(x)\) to any literal due to \(\sigma\). As
we are stuck at this point, we mark this with \Lightning.
\autoref{fig:clausal-calc-ex} shows a derivation for \(M\) that
corresponds to that proof search. The extension steps in the derivation
are labelled like the corresponding proof steps in matrix (1). The
derivation is \emph{not} a connection proof for \(M\), because the leaf
\(\langle \left\{\mathsf{q}(x)\right\}, M, \left\{\right\} \rangle\) is
not an axiom.\end{example}

\begin{figure}
\includegraphics{prftree/clausal-calc-ex.tex}
\caption{Derivation with \(\sigma(x) = \sigma(y) = \mathsf{a}\).}
\label{fig:clausal-calc-ex}
\end{figure}

This example illustrates that search in connection tableaux is not
confluent, i.e., we can end up with unprovable leaves in derivations for
a matrix \(M\) although the formula corresponding to \(M\) is
unsatisfiable. This makes it necessary to backtrack to previous states
of derivations to obtain a complete proof search method. We will study
two backtracking strategies in the next section.

\hypertarget{backtracking}{%
\section{Backtracking}\label{backtracking}}

In the failed proof attempt shown in \autoref{ex:failed}, we frequently
talked about \emph{obligations} and \emph{choices}. Fulfilling
obligations assures soundness, and making alternative choices
exhaustively assures completeness.

Otten's \emph{unrestricted backtracking} strategy
\citep{DBLP:conf/cade/Otten08} is sound and complete. It makes choices
until an obligation cannot be fulfilled. At this point, the strategy
changes the most recent choice for which there is an untried
alternative. The strategy succeeds if it fulfills all obligations, and
fails if it runs out of alternatives.

\begin{example}[Unrestricted Backtracking]\label{ex:complete}Consider
the proof search attempt in \autoref{ex:failed}. The last choice we made
in that example was to connect \(\mathsf{r}(y)\) to
\(\lnot \mathsf{r}(\mathsf{a})\) as part of step (1.2). We can make a
different choice here, namely connect \(\mathsf{r}(y)\) to
\(\lnot \mathsf{r}(\mathsf{b})\), which we perform in step (2.1).
However, as it turns out, this will not help us once we have to deal
with \(\mathsf{q}(x)\) anew in step (2.2), for now we have
\(\sigma(x) = \sigma(y) = \mathsf{b}\), which still does not permit a
connection from \(\mathsf{q}(x)\). So we backtrack again, leading to the
proof search shown in matrix (3). This time, the last choice was to
connect \(\mathsf{r}(y)\) to \(\lnot \mathsf{r}(\mathsf{b})\), but now,
we cannot find a different way to connect \(\mathsf{r}(y)\). So we look
at our second to last choice, namely to connect \(\mathsf{p}(x)\) to
\(\lnot \mathsf{p}(y)\). We can make an alternative choice here as step
(3.1), namely to connect \(\mathsf{p}(x)\) to \(\lnot \mathsf{p}(z)\).
Now we are back once more to the dreaded \(\mathsf{q}(x)\), but finally,
due to \(\sigma(x) = \sigma(z)\) not pointing to an actual term, we can
connect \(\mathsf{q}(x)\) to \(\lnot \mathsf{q}(\mathsf{c})\) as step
(3.2). This concludes the proof, as we have no more obligations left at
this point.\end{example}

Otten's \emph{restricted backtracking} strategy
\citep{DBLP:journals/aicom/Otten10} is sound, but incomplete. However,
it is often significantly more effective than the complete strategy. To
define it, Otten introduces the property of \emph{solvedness} on
literals in a proof search.

\begin{definition}[Principal literal, solved
literal]\label{def:solves}When the reduction or extension rules are
applied, the literal \(L\) (see \autoref{fig:clausal-calculus}) is
called the \emph{principal literal} of the proof step. A reduction step
\emph{solves} a literal \(L\) iff \(L\) is its principal literal. An
extension step \(S\) \emph{solves} a literal \(L\) iff \(L\) is the
principal literal of \(S\) and there is a proof for the left premise of
\(S\), i.e., there is a derivation for the left premise of \(S\) having
only axioms as leaves.\end{definition}

The restricted backtracking strategy works like the unrestricted one,
with one exception: Once a literal is solved, restricted backtracking
discards all choices to solve the literal differently.

\begin{example}[Restricted Backtracking]\label{ex:restricted}Consider
matrix (1) of \autoref{fig:search}. Proof step (1.1) does not solve any
literal, so at this point, proof search behaves like in
\autoref{ex:complete}. Proof step (1.2) solves \(\mathsf{r}(y)\), so at
that point, alternative choices to solve \(\mathsf{r}(y)\) are
discarded. At the same time, step (1.2) also solves \(\mathsf{p}(x)\),
so alternative choices to solve \(\mathsf{p}(x)\) are discarded as well.
In proof step (1.3), we note that \(\mathsf{q}(x)\) cannot be connected.
However, unlike in \autoref{ex:complete}, we have no alternative choices
left to backtrack to because they were discarded as a result of step
(1.2). That means that the restricted backtracking strategy cannot find
a proof once we commit to connecting \(\mathsf{p}(x)\) to
\(\lnot \mathsf{p}(y)\) as first step.\end{example}

\hypertarget{less-restricted}{%
\section{Less Restricted Backtracking}\label{less-restricted}}

Restricted backtracking can be decomposed into two cuts: cuts on
reduction and cuts on extension steps. Kaliszyk already implemented
these two cuts separately, but did not describe it, as they are usually
most useful in conjunction. Here, the distinction arises naturally, as
it allows to more succinctly describe a new backtracking strategy.

I will now distinguish \emph{inclusive} and \emph{exclusive} cuts. An
inclusive cut discards all alternatives to solve a literal, whereas an
exclusive cut discards all alternatives to solve a literal, \emph{except
for derivations starting with a different proof step}. Otten's
restricted backtracking strategy shown in \autoref{backtracking} uses
inclusive cuts on both reduction and extension steps. To the best of my
knowledge, exclusive cuts have not been researched before.

\begin{example}[Exclusive Cut]\label{ex:exclusive}We will, for the last
time, revisit the proof search in \autoref{fig:search}. After proof step
(1.2), exclusive cut discards alternative ways to solve
\(\mathsf{p}(x)\), except for derivations starting with different
extension steps. As a result, after being stuck at step (1.3) with
\(\mathsf{q}(x)\), we \emph{can} backtrack unlike in
\autoref{ex:restricted}, namely to the proof search in matrix (3),
because it solves \(\mathsf{p}(x)\) in step (3.1) starting with a
different extension step. From there, proof search behaves again like in
\autoref{ex:complete}, solving the problem in step (3.2) after
connecting \(\mathsf{q}(x)\).\end{example}

To sum up the outcomes of different backtracking strategies on the proof
search in \autoref{fig:search}: The complete strategy
(\autoref{ex:complete}) solves the problem, going through all stages
from (1) to (3). The exclusive cut (\autoref{ex:exclusive}) also solves
the problem, but takes one stage less, going through only (1) and (3).
The inclusive cut (\autoref{ex:restricted}) fails after (1). For this
example, exclusive cut therefore is the most efficient strategy.

For reduction steps, an exclusive cut is equivalent to no cut, so I
distinguish between inclusive and exclusive cut only for extension
steps. I abbreviate (inclusive) cut on reduction steps as R and
inclusive and exclusive cut on extension steps as EI and EX,
respectively. Otten's restricted backtracking strategy can be described
as a combination of R and EI, written as REI.

\begin{figure}
\includegraphics{tikz/cuts.tex}
\caption{Effect of different cuts on the tree of alternatives.}
\label{fig:cuts}
\end{figure}

\autoref{fig:cuts} visualises the alternatives that are cut once a
literal is solved. In each of the trees, the left child of the root is a
proof step \(S\) that solves a literal, the children of the left child
are alternatives to proof steps that are descendants of \(S\), and the
right child is the alternative to \(S\). Reduction and extension steps
are marked as R and E, respectively, and alternatives are marked as
``?''. Both the R and EI cut are inclusive cuts because the right child
is cut, and the EX cut is exclusive because the right child is
preserved. Both EI and EX cuts eliminate all alternatives below the
proof step.

The seemingly small difference between inclusive and exclusive cut has a
large impact on the effectiveness of the prover. We will see this in the
evaluation (\autoref{evaluation}).

\hypertarget{implementation}{%
\section{Implementation}\label{implementation}}

The connection prover leanCoP is compactly implemented in Prolog as a
recursive predicate \texttt{prove} that takes \(C\) and \(Path\) as
parameters, using a helper predicate \texttt{lit} that models \(M\)
\citep{DBLP:conf/cade/Otten08}. leanCoP implements restricted
backtracking using Prolog's built-in cut operator
\citep{DBLP:journals/aicom/Otten10}. Kaliszyk has reimplemented leanCoP
using stacks for backtracking \citep{DBLP:conf/tableaux/Kaliszyk15}.

Based on Kaliszyk's stack-based implementation, I implemented a
connection prover called \emph{meanCoP} in Rust.\footnote{The name
  meanCoP abbreviates ``more efficient, albeit non-lean connection
  prover''. The source code of meanCoP is available at
  \url{https://github.com/01mf02/cop-rs}. I evaluated revision 884aea4
  compiled with Rust 1.49.} Like C++, Rust favours zero-overhead
abstractions \citep{DBLP:conf/esop/Stroustrup12}, making it a suitable
candidate for the development of high-performance automated theorem
provers. I use functional programming for preprocessing and imperative
programming for the prover loop. I follow Kaliszyk's implementation for
the prover loop, but I use dynamic instead of static arrays in order to
allow for arbitrarily sized stacks, terms, etc. The prover loop does not
use Rust's standard library and can be therefore compiled to targets
such as WASM, which can be used to create websites with an embedded
prover that is run locally in a web browser. Furthermore, meanCoP
contains a tiny proof checker that is run before outputting a proof.
This is useful to assure that the prover is sound.

meanCoP supports most major features of leanCoP, such as
conjecture-directed search, regularity, and lemmas
\citep{DBLP:conf/cade/Otten08}. Furthermore, meanCoP supports the R, EI,
and EX cuts (\autoref{less-restricted}). By default, meanCoP uses
(inclusive) cut on lemma steps, which does not hamper completeness, as
lemma steps do not impact the substitution.

\begin{figure}
\includegraphics{tikz/stack.tex}
\caption{Effect of different cuts on the stack of alternatives.}
\label{fig:stack}
\end{figure}

Kaliszyk uses a stack of alternatives to keep track of proof steps to
backtrack to. This allows for a compact implementation of inclusive and
exclusive cuts. \autoref{fig:stack} shows the effect of inclusive and
exclusive cut on the stack of alternatives. \autoref{fig:nocut} shows
the initial situation of the stack after a literal was solved with a
proof step whose alternative is \(a_n\). Above \(a_n\) are alternatives
to proof steps added after \(a_n\), and below \(a_n\) are alternatives
to proof steps added before \(a_n\). Using no cut does not change the
stack at this point. Both exclusive and inclusive cut eliminate all
alternatives added after \(a_n\), but the exclusive cut
(\autoref{fig:cutexclusive}) keeps one alternative more than the
inclusive cut (\autoref{fig:cutinclusive}), namely \(a_n\). Using
inclusive instead of exclusive cut amounts to truncating the stack of
alternatives to length \(n\) instead of length \(n-1\) once a literal
was solved.

\hypertarget{evaluation}{%
\section{Evaluation}\label{evaluation}}

I evaluate the performance of meanCoP (\autoref{implementation}) and
other provers on several first-order problem datasets.\footnote{The
  evaluation results are available in more detail at
  \url{http://cl-informatik.uibk.ac.at/~mfaerber/arecca-2023.html}.} For
every dataset and prover, I measure the number of problems solved by the
prover in a given time. All evaluated connection provers use a single
strategy with conjecture-directed search and non-definitional
(i.e.~standard or naive) translation into CNF
\citep{DBLP:journals/jsc/PlaistedG86, DBLP:journals/aicom/Otten10},
unless specified otherwise. I use the same hardware, the same timeout,
and the same datasets as in my previous evaluation of connection provers
together with Kaliszyk and Urban \citep{DBLP:journals/jar/FarberKU21}. I
will compare the results in this paper with those of the previous
evaluation.

I use a 48-core server with AMD Opteron 6174 2.2GHz CPUs, 320 GB RAM,
and 0.5 MB L2 cache per CPU. Each problem is always assigned one CPU. I
run every prover with a timeout of 10 seconds per problem.

I use several first-order logic datasets for evaluation, with statistics
given in \autoref{tab:fof-datasets}:

\begin{itemize}
\tightlist
\item
  TPTP \citep{DBLP:journals/jar/Sutcliffe17} is a large benchmark for
  automated theorem provers. It is used in CASC
  \citep{DBLP:journals/aicom/Sutcliffe16}. Its problems are based on
  different logics and come from various domains. I use the nonclausal
  first-order problems (files matching \texttt{*+?.p}) of TPTP~6.3.0.
\item
  MPTP2078 \citep{DBLP:journals/jar/AlamaHKTU14} contains 2078 problems
  exported from the Mizar Mathematical Library. It comes in the two
  flavours ``bushy'' and ``chainy'': In the ``chainy'' dataset, every
  problem contains all facts stated before the problem, whereas in the
  ``bushy'' dataset, every problem contains only the Mizar premises
  required to prove the problem.
\item
  Miz40 contains the problems from the Mizar library for which at least
  one ATP proof has been found using one of the 14 combinations of
  provers and premise selection methods considered in
  \citep{DBLP:journals/jar/KaliszykU15a}. The problems are translated to
  untyped first-order logic using the MPTP infrastructure
  \citep{DBLP:journals/jar/Urban04}. The problems are minimised using
  ATP-based minimisation, i.e., re-running the ATP only with the set of
  proof-needed axioms until this set no longer becomes smaller. This
  typically leads to even better axiom pruning and ATP-easier problems
  than in the Mizar-based pruning used for the ``bushy'' version above.
\item
  FS-top is a translation to first-order logic of the top-level HOL
  Light theorems of the Flyspeck project, which finished in 2014 a
  formal proof of the Kepler conjecture \citep{Hales2017}.
\end{itemize}

\begin{table}

\caption{Evaluation datasets and the number of contained first-order
problems. \label{tab:fof-datasets}}

\begin{tabular}{lrrrrr}

\toprule

Dataset & TPTP & bushy & chainy & Miz40 & FS-top\tabularnewline

\midrule

Problems & 7492 & 2078 & 2078 & 32524 & 27111\tabularnewline

\bottomrule

\end{tabular}

\end{table}

\hypertarget{comparison-of-meancop-strategies}{%
\subsection{Comparison of meanCoP
strategies}\label{comparison-of-meancop-strategies}}

The first part of the evaluation studies the impact of different
combinations of cuts on the number of problems solved by meanCoP.

\begin{table}

\caption{Number of solved problems. \label{tab:evaluation}}

\begin{tabular}{lrrrrr}

\toprule

Cut & TPTP & bushy & chainy & Miz40 & FS-top\tabularnewline

\midrule

None & 1731 & 546 & 208 & 9247 & 4038\tabularnewline
R & 1857 & 644 & 252 & 12965 & 4447\tabularnewline
EI & 1984 & 724 & 333 & 13853 & 4249\tabularnewline
EX & 2056 & 820 & 268 & 15507 & 4758\tabularnewline
REI & 1988 & 730 & \textbf{341} & 13562 & 4267\tabularnewline
REX & \textbf{2126} & \textbf{850} & 294 & \textbf{16135} & \textbf{4994}\tabularnewline

\bottomrule

\end{tabular}

\end{table}

\autoref{tab:evaluation} shows the number of problems solved by
strategy. For all datasets, the complete strategy without any cut solves
the least problems. Among the previously implemented cuts, namely R, EI,
and REI, REI solves the most problems, except on the dataset FS-top,
where R prevails. Adding cut on reduction (R) to any strategy increases
the number of solved problems, except for the Miz40 dataset. The
strategies with exclusive cut on extension steps (EX, REX) outperform
those with inclusive cut (EI, REI) on all datasets except for the chainy
one. I explore the reason for this in \autoref{proof-analysis}.

On most datasets, the strategies using exclusive cut bring an impressive
improvement of the prover power. The REX strategy increases the number
of solved problems compared to the REI strategy by 16.4\% for bushy,
17.0\% for FS-top (12.3\% if we compare with the R cut), and 19.0\% for
Miz40. Remarkably, on TPTP, the improvement turns out much smaller with
only 6.9\%.

\begin{table}

\caption{Union of solved problems. \label{tab:union}}

\begin{tabular}{lrrrrr}

\toprule

Cut & TPTP & bushy & chainy & Miz40 & FS-top\tabularnewline

\midrule

Any but (R)EX & 2277 & 834 & 391 & 15885 & 5185\tabularnewline
REX and REI & 2297 & 907 & 389 & 16803 & 5191\tabularnewline
Any & 2387 & 927 & 402 & 17437 & 5560\tabularnewline

\bottomrule

\end{tabular}

\end{table}

\autoref{tab:union} shows the union of problems solved by a portfolio of
strategies. The first row shows the problems solved by any of the four
previously used cut strategies, including the unrestricted backtracking
strategy without cut, but also combinations of cut on reduction and
inclusive cut on extension, while excluding our new exclusive cut.
Comparing the first row with the REX results from
\autoref{tab:evaluation}, we see that the new REX strategy solves
single-handedly more problems than a union of four strategies on the
bushy and Miz40 datasets, which is quite noteworthy. The second row
shows the problems solved by any of the two most powerful strategies,
including the REX strategy that uses exclusive cut. This combination is
better than the combination of all previous cut strategies in the first
row on all datasets except for chainy, where it is only two problems
behind. Combining all strategies (row 3) clearly boosts the number of
solved problems compared to the previously available strategies (row 1),
namely 11.2\% for bushy, 4.8\% for TPTP, 9.8\% for Miz40, and 7.2\% for
FS-top.

In conclusion, the new strategies do not only prove more problems, but
the problems they solve are also sufficiently complementary from the
problems solved by previously available strategies. This makes the new
strategies attractive in portfolio modes.

\hypertarget{proof-analysis}{%
\subsection{Proof analysis}\label{proof-analysis}}

I compare the proofs of the complete, REX, and REI strategies, similar
to Otten's comparison of the complete and REI strategies
\citep[sec.~4.2]{DBLP:journals/aicom/Otten10}.

There are two indicators for the quality of a cut strategy C2 with
respect to a more complete cut strategy C1: which percentage of C1
proofs C2 finds, and how many more inferences C1 takes to find these
proofs. When two problems are solved identically by C2 and C1, the
additional backtracking done by C1 is superfluous for the proof. The
fewer inferences C2 takes to find identical proofs, the more likely it
is that C2 also finds proofs which are out of reach for C1 in a given
time limit.

\autoref{tab:percentage-proofs} shows the number of problems for which
two strategies find identical proofs. To understand how these numbers
emerge, let us consider the chainy dataset, comparing the complete (C1 =
None) and the REX (C2 = REX) strategies. Here, \autoref{tab:evaluation}
shows us that C1 solves 208 and C2 solves 294 chainy problems. C2 finds
for 186 of the 208 problems solved by C1 the same proof as C1, which
amounts to the 89.4\% given in \autoref{tab:percentage-proofs}. Note
that C2 solves 203 of the 208 problems that C1 solves, which means that
for 17 problems, it finds different proofs than C1.

Of all proofs found by the complete strategy, REX finds between 89.4\%
(chainy) to 66.5\% (bushy), whereas REI finds only between 68.3\% (TPTP)
and 46.7\% (bushy). REI also finds only between 66.6\% (FS-top) and
40.8\% (bushy) of the proofs found by REX. This shows that there are
significantly fewer proofs requiring unrestricted backtracking (no cut)
than proofs requiring backtracking that replaces root steps (REX).

\begin{table}

\caption{Percentage of problems solved by C1 that are identically solved
by C2. \label{tab:percentage-proofs}}

\begin{tabular}{llrrrrr}

\toprule

C1 & C2 & TPTP & bushy & chainy & Miz40 & FS-top\tabularnewline

\midrule

None & REX & 84.5 & 66.5 & 89.4 & 77.8 & 81.0\tabularnewline
None & REI & 68.3 & 46.7 & 57.7 & 54.2 & 67.4\tabularnewline
REX & REI & 63.3 & 40.8 & 59.2 & 50.1 & 66.6\tabularnewline

\bottomrule

\end{tabular}

\end{table}

\begin{table}

\caption{Ratio between sum of inferences taken by C1 and inferences
taken by C2, for problems identically solved by C1 and C2.
\label{tab:ratio-infs}}

\begin{tabular}{llrrrrr}

\toprule

C1 & C2 & TPTP & bushy & chainy & Miz40 & FS-top\tabularnewline

\midrule

None & REX & 4.4 & 37.0 & 9.9 & 37.4 & 19.8\tabularnewline
None & REI & 4.2 & 55.4 & 32.0 & 54.6 & 28.8\tabularnewline
REX & REI & 3.3 & 4.0 & \textbf{8.4} & 2.4 & 2.2\tabularnewline

\bottomrule

\end{tabular}

\end{table}

\begin{table}

\caption{Average of ratios between inferences taken by C1 and inferences
taken by C2, for problems identically solved by C1 and C2.
\label{tab:avg-ratio-infs}}

\begin{tabular}{llrrrrr}

\toprule

C1 & C2 & TPTP & bushy & chainy & Miz40 & FS-top\tabularnewline

\midrule

None & REX & 7.3 & 22.1 & 5.9 & 62.8 & 12.5\tabularnewline
None & REI & 41.5 & 28.1 & 23.7 & 78.9 & 32.0\tabularnewline
REX & REI & \textbf{39.6} & 2.9 & \textbf{6.9} & 3.7 & 2.8\tabularnewline

\bottomrule

\end{tabular}

\end{table}

We are now going to analyse how much different strategies reduce the
search space. For this, we will compare the number of inferences taken
by two strategies when they find the same proofs. Given two strategies
C1 and C2, we can construct an \(I\) as follows: if C1 and C2 found the
same proof \(p\) for a problem, then \((p, n_1, n_2) \in I\), where
\(n_1\) and \(n_2\) are the number of inferences taken by C1 and C2.
\autoref{tab:ratio-infs} shows the ratio of the sum of all inferences,
calculated by
\[\frac{\sum _{(p, n_1, n_2) \in I} n_1}{\sum _{(p, n_1, n_2) \in I} n_2},\]
and \autoref{tab:avg-ratio-infs} shows the average of the ratios of
inferences, calculated by
\[\sum _{(p, n_1, n_2) \in I} \frac{n_1}{n_2} \div \lvert I \rvert.\] In
both cases, the higher a value, the more C2 reduces the search space
with respect to C1.

Let us look at \autoref{tab:ratio-infs}. For example, on the Miz40
dataset, for all problems identically solved by REX and the complete
strategy, the sum of inferences by the complete strategy is 37.4 times
the sum of inferences by the REX strategy. This is the highest ratio for
REX with respect to the complete strategy. REX also achieves on Miz40
the largest increase of solved problems compared to the complete
strategy (+74.5\%). Conversely, on TPTP, where REX shows the smallest
inference ratio (4.4), REX also least improves the number of solved
problems (+22.8\%). On most datasets, the ratios between REI and REX are
significantly smaller than the ratios between REX/REI and the complete
strategy; for example, on the Miz40 dataset, REI reduces inferences
compared to REX only by 2.4, whereas REX and REI reduce inferences
compared to the complete strategy by 37.4 and 54.6. This indicates that
REX and REI are much ``closer'' to each other than to the complete
strategy. Notable exceptions are the chainy and TPTP datasets, where the
ratios between the complete strategy and REX are quite similar to the
ratios between REX and REI.\footnote{This can be also seen in
  \autoref{tab:avg-ratio-infs}, where the ratio between REX and REI on
  TPTP is unusually high and clearly exceeds the ratio between REX and
  the complete strategy.} Interestingly, these are the datasets where
REX proves fewer (chainy) or only few more (TPTP) problems than REI. On
datasets like chainy that contain many problems with unusually many
axioms, implying a larger explosion of the search space, a more
aggressive cut such as REI turns out to be beneficial. In general, we
can observe that REX yields the best results when REX greatly reduces
the search space with respect to the complete strategy and REI slightly
reduces the search space with respect to REX.

In summary, REX is successful because it conserves a considerable amount
of existing proofs, while sufficiently reducing the number of inferences
in order to find new proofs.

\hypertarget{comparison-with-other-leancop-implementations}{%
\subsection{Comparison with other leanCoP
implementations}\label{comparison-with-other-leancop-implementations}}

I evaluate meanCoP and two other implementations of leanCoP, namely
leanCoP 2.1 using the Prolog compiler ECLiPSe 5.10, and fleanCoP, which
is a reimplementation of leanCoP in OCaml using streams
\citep{DBLP:journals/jar/FarberKU21}. All evaluated connection provers
in this section use a single strategy with conjecture-directed search,
non-definitional translation into CNF, and restricted backtracking,
i.e.~REI.\footnote{This amounts to running meanCoP with
  \texttt{-\/-conj\ -\/-cuts\ rei}, leanCoP with
  \texttt{SET=\textquotesingle{}{[}nodef,conj,cut{]}\textquotesingle{}},
  and fleanCoP with \texttt{-schedule\ 0\ -nodefcnf}.} Care is taken
that leanCoP-REI, fleanCoP-REI, and meanCoP-REI perform the same
inferences.

\begin{table}

\caption{Prover runtime in seconds for problems solved by leanCoP-REI.
\label{tab:runtime}}

\begin{tabular}{rrrrrr}

\toprule

Prover & TPTP & bushy & chainy & Miz40 & FS-top\tabularnewline

\midrule

leanCoP-REI & 1299.7 & 461.9 & 319.1 & 9308.7 & 2451.6\tabularnewline
fleanCoP-REI & 488.1 & 190.9 & 69.8 & 3845.6 & 657.2\tabularnewline
meanCoP-REI & 200.0 & 17.3 & 29.0 & 347.9 & 88.5\tabularnewline

\bottomrule

\end{tabular}

\end{table}

\autoref{tab:runtime} shows the runtime of different leanCoP
implementations on sets of problems solved by the original leanCoP. The
meanCoP prover is between 27.7 (FS-top) and 6.5 (TPTP) times faster than
the original leanCoP and between 11.1 (Miz40) and 2.4 (TPTP) times
faster than its OCaml reimplementation using streams.

\begin{table}

\caption{Number of solved problems for different leanCoP
implementations. \label{tab:solved-lc}}

\begin{tabular}{rrrrrr}

\toprule

Prover & TPTP & bushy & chainy & Miz40 & FS-top\tabularnewline

\midrule

leanCoP-REI & 1673 & 606 & 182 & 11243 & 3664\tabularnewline
fleanCoP-REI & 1859 & 670 & 289 & 12204 & 3980\tabularnewline
meanCoP-REI & 1988 & 730 & 341 & 13562 & 4267\tabularnewline

\bottomrule

\end{tabular}

\end{table}

\autoref{tab:solved-lc} shows that the higher performance of meanCoP
translates to a vastly increased number of proven problems. The largest
improvement can be seen on the chainy dataset, where leanCoP, fleanCoP
and meanCoP (all using the REI strategy) prove 182, 289 (+58.8\%), and
341 (+87.4\% compared to leanCoP and +18.0\% compared to fleanCoP)
problems, respectively.

\hypertarget{comparison-with-other-provers}{%
\subsection{Comparison with other
provers}\label{comparison-with-other-provers}}

I compare meanCoP with several non-connection provers that I previously
evaluated in joint work with Kaliszyk and Urban \citep[table
2]{DBLP:journals/jar/FarberKU21}. In particular, I evaluate Vampire 4.0
\citep{DBLP:conf/cav/KovacsV13} and E 2.0
\citep{DBLP:conf/lpar/Schulz13}, which performed best in the first-order
category of CASC-J8 \citep{DBLP:journals/aicom/Sutcliffe16}. Vampire and
E are written in C++ and C, respectively, implement the superposition
calculus, and perform premise selection with SInE
\citep{DBLP:conf/cade/HoderV11}. Furthermore, Vampire integrates several
SAT solvers \citep{DBLP:conf/micai/BiereDKV14}, and E automatically
determines proof search settings for a given problem. I run E with
\texttt{-\/-auto-schedule} and Vampire with \texttt{-\/-mode\ casc}. In
addition, I evaluate the ATP Metis 2.3 (release 20171005)
\citep{Hurd03}: It implements the ordered paramodulation calculus
(having inference rules for equality just like the superposition
calculus), but is considerably smaller than Vampire and E and is
implemented in Standard ML.

I also evaluate two versions of leanCoP 2.1: First, I evaluate leanCoP
2.1 with strategy scheduling, which will be simply called ``leanCoP'' in
this section. Running leanCoP with a timeout of 10 seconds runs about 10
different search strategies for one second each. Second, I evaluate the
first strategy in the strategy schedule of leanCoP 2.1 which searches
using restricted backtracking until a path limit of 7, then switches to
a complete search with unrestricted backtracking. I call this strategy
``leanCoP-CC7''. Unlike all other evaluated connection provers with a
single strategy, leanCoP-CC7 does not use conjecture-directed search;
furthermore, it uses definitional translation into CNF for the
conjecture of the input problem. Finally, I evaluate an adapted version
of leanCoP-CC7 with conjecture-directed search and with non-definitional
translation into CNF. I call this strategy ``leanCoP-NCCC7''.\footnote{leanCoP-CC7
  and leanCoP-NCCC7 amount to running leanCoP with
  \texttt{SET=\textquotesingle{}{[}cut,comp(7){]}\textquotesingle{}} and
  \texttt{SET=\textquotesingle{}{[}nodef,conj,cut,comp(7){]}\textquotesingle{}},
  respectively.}

\begin{table}

\caption{Number of solved problems by different provers.
\label{tab:provers}}

\begin{tabular}{lrrrrr}

\toprule

Prover & TPTP & bushy & chainy & Miz40 & FS-top\tabularnewline

\midrule

Vampire & 4404 & 1253 & 656 & 30341 & 6358\tabularnewline
E & 3664 & 1167 & 287 & 26003 & 7382\tabularnewline
Metis & 1376 & 500 & 75 & 18519 & 3537\tabularnewline
leanCoP-CC7 & 1749 & 635 & 154 & 13121 & 3892\tabularnewline
leanCoP-NCCC7 & 1752 & 651 & 188 & 11637 & 4188\tabularnewline
leanCoP & 1917 & 673 & 196 & 13636 & 4373\tabularnewline
meanCoP-REI & 1988 & 730 & \textbf{341} & 13562 & 4267\tabularnewline
meanCoP-REX & 2126 & 850 & 294 & 16135 & 4994\tabularnewline

\bottomrule

\end{tabular}

\end{table}

\begin{table}

\caption{Number of problems solved by meanCoP-REX, but not by another
prover. \label{tab:provers-norex}}

\begin{tabular}{lrrrrr}

\toprule

Prover & TPTP & bushy & chainy & Miz40 & FS-top\tabularnewline

\midrule

Vampire & 33 & 19 & 21 & 88 & 1038\tabularnewline
E & 190 & 65 & 92 & 1803 & 720\tabularnewline
Metis & 972 & 390 & 222 & 3111 & 2463\tabularnewline
leanCoP & 383 & 207 & 104 & 3394 & 908\tabularnewline
Any above & 15 & 12 & 16 & 15 & 176\tabularnewline

\bottomrule

\end{tabular}

\end{table}

\autoref{tab:provers} shows the results: Vampire proves most problems on
all datasets except for FS-top, where E prevails. On the chainy dataset,
meanCoP proves more problems than E, which is likely due to the
conjecture-directed search. Metis proves the fewest problems, except on
the Miz40 dataset, where it proves more problems than any connection
prover, but less than Vampire and E. leanCoP-CC7 proves more problems
than Metis on all datasets except for Miz40, but proves fewer problems
than leanCoP (with strategy scheduling) on all datasets. meanCoP-REI
proves more problems than leanCoP on all datasets but Miz40 and FS-top,
and meanCoP-REX proves more problems than leanCoP on all datasets.

\autoref{tab:provers-norex} shows for several provers \(P\) how many
problems meanCoP-REX can solve that were not solved by \(P\). For
example, it shows that meanCoP-REX proves 1038 FS-top problems that were
not solved by Vampire, which solves 6358 problems in total. The last
line in \autoref{tab:provers-norex} shows the number of problems that
meanCoP-REX solves which no other prover in the table solves.

\begin{figure}
\includegraphics{tikz/plot.tex}
\caption{Evolution of number of solved bushy problems over time.}
\label{fig:plot}
\end{figure}

\autoref{fig:plot} shows for several provers the number of problems on
the bushy dataset proved up to a certain time. meanCoP-REX proves
considerably more problems than Vampire in the first 10 milliseconds,
namely 428 versus a single one. However, Vampire catches up after about
50 milliseconds, leaving all other provers behind. leanCoP and
leanCoP-REI solve their first problem about 50 milliseconds after any
other prover, which is due to the relatively high start-up time caused
by the compilation of the prover at each run. After 50 milliseconds, the
order between the provers remains stable. The curves for the provers
without strategy scheduling (leanCoP-REI, meanCoP-REI, meanCoP-REX)
flatten with time, whereas the curves for Vampire and leanCoP shows
several ``bumps'' due to strategy scheduling. The average time used to
solve a problem is 0.53 seconds for meanCoP-REI, 0.64 seconds for
meanCoP-REX, 0.76 seconds for leanCoP-REI, 0.87 seconds for leanCoP, and
0.99 seconds for Vampire.

\hypertarget{related-work}{%
\section{Related Work}\label{related-work}}

The MaLeCoP prover by Urban et al.~\citep{DBLP:conf/tableaux/UrbanVS11}
and the FEMaLeCoP prover by Kaliszyk and Urban
\citep{DBLP:conf/lpar/KaliszykU15} were among the first to use machine
learning to guide connection proof search. These provers order the
applicable extension steps in prover states by Naive Bayesian
probabilities that are inferred from previous proofs. Like leanCoP, they
use depth-first search, iterative deepening, and backtracking, which
makes such provers likely to benefit from advances in backtracking
strategies as presented in this work.

Other connection provers have moved away more from leanCoP's traditional
backtracking-based search. I developed monteCoP in joint work with
Kaliszyk and Urban \citep{DBLP:journals/jar/FarberKU21}, Kaliszyk et
al.~developed rlCoP \citep{DBLP:conf/nips/KaliszykUMO18}, and Olšák et
al.~developed follow-up work to rlCoP \citep{DBLP:conf/ecai/OlsakKU20}.
All these provers use machine-learnt policies to explore the search
space, with Monte Carlo Tree Search taking the role that backtracking
plays in leanCoP. For that reason, such provers can probably not
directly profit from this work.

We evaluate FEMaLeCoP and monteCoP on the bushy dataset, using 60
seconds timeout, definitional clausification and the REI strategy.
Comparing the non-learning with the learning versions of the provers,
the increase in number of solved problems is from 563 to 601 for
monteCoP (+6.7\%) and from 577 to 592 for FEMaLeCoP (+2.6\%)
\citep[table 8]{DBLP:journals/jar/FarberKU21}, thus far below the
increase of 16.4\% gained in the current work.

Kaliszyk et al.~evaluate rlCoP on the Miz40 dataset, where it proves
16108 problems after 10 iterations of training. Although I evaluate
meanCoP on the same dataset, where meanCoP proves 16134 problems, it is
unfortunately difficult to compare the results for two reasons: First,
Kaliszyk et al.~limit the number of inferences instead of the time
allotted to the prover. Second, most inferences performed by rlCoP end
up in prover states that are not actually explored, due to not being
chosen by Monte Carlo Tree Search.

Another line of work extends connection provers with native support for
equality. Rawson's lazyCoP is a connection prover based on Paskevich's
connection tableaux calculus with lazy paramodulation
\citep{Sutcliffe20, DBLP:journals/jar/Paskevich08}. It supports
first-order logic with equality. Given that lazyCoP does not use
backtracking to control the search, it seems unlikely that exclusive cut
could be integrated in this system.

Otten's ileanCoP for intuitionistic logic \citep{DBLP:conf/cade/Otten08}
and MleanCoP for modal logic \citep{DBLP:conf/cade/Otten14}, as well as
nanoCoP for nonclausal proof search \citep{DBLP:conf/ijcai/Otten17},
could all integrate exclusive cut seamlessly.

\hypertarget{conclusion}{%
\section{Conclusion}\label{conclusion}}

I introduced a new kind of cut on extension steps called exclusive cut,
which discards all alternatives to solve a literal, except for
derivations starting with a different extension step. I implemented the
described techniques in a new prover called meanCoP. Evaluating meanCoP
on several first-order problem datasets yielded that a combination of
cut on reduction steps and exclusive cut on extension steps (REX)
improves the number of solved problems compared to the previous best
strategy by up to 19\%.

\hypertarget{acknowledgements}{%
\subsubsection*{Acknowledgements}\label{acknowledgements}}
\addcontentsline{toc}{subsubsection}{Acknowledgements}

I am grateful to the anonymous CADE, TABLEAUX, and AReCCa reviewers as
well as to Jasmin Blanchette, Mathias Fleury, Cezary Kaliszyk, and Petar
Vukmirović for their comments on drafts of this paper. This work has
been supported by the Schrödinger grant (J~4386) of the Austrian Science
Fund (FWF).

\bibliography{literature}

\end{document}